\documentstyle[12pt]{article}
\title{Haldane gap in a S=1 exchange model with long-range interactions}
\author{P.D. Sacramento$^{1,*}$ and V.R. Vieira$^{2}$ \\ 
$^{1}$Department of Physics and Astronomy \\
The Johns Hopkins University, Baltimore, MD 21218, USA \\
$^{2}$Centro de F\'{\i}sica das Interac\c{c}\~{o}es Fundamentais \\
Instituto Superior T\'{e}cnico, 
Av. Rovisco Pais, 1096 Lisboa Codex, Portugal} 
\date{}

\begin{document}
\maketitle
\begin{abstract}
The ground-state properties of the $S=1$ Haldane-Shastry model are
studied
using a modified Lanczos algorithm and 
diagonalizing exactly small chains. We find evidence that,
as for the antiferromagnetic Heisenberg model, the spectrum shows
a gap, in contrast to the $S=\frac{1}{2}$ case. The correlation
functions $<S^{z}(0)S^{z}(m)>$ decay exponentially for large $m$. We find that 
the correlation functions for the Haldane-Shastry model decay
faster than for the Heisenberg model. We estimate the infinite system limit 
for the ground-state energy, value of the gap and correlation functions.
\end{abstract}

The $S=\frac{1}{2}$ Haldane-Shastry model [1,2] has attracted considerable
attention since it is an integrable model that belongs to
a class of systems with long-range interactions. The ground-state
energy and correlation functions have been obtained [1-3] together
with the thermodynamics [4]. The ground-state wave function is a
spin singlet of the Jastrow-Gutzwiller form. The excitations are
spin-$\frac{1}{2}$ spinons [4] that form a gas of a semionic nature [4,5].
The asymptotic correlations decay algebraically with exponent $\eta =1$
without logarithmic corrections, in contrast to the Heisenberg case. This
indicates the absence of spin exchange between the spinons rendering the
model solvable in greater detail than the short-range Heisenberg
counterpart, solvable by the traditional Bethe ansatz method.

On the other hand, the similarities to the Heisenberg model lead to
the interest in studying the model for higher values of the spin. In 
particular, the case of spin $S=1$ (and in general the integer spin
cases) are of interest since it has been established on general grounds
and verified explicitly for short-range interactions,
that these values are qualitatively different from the half-integer ones
due to the Haldane gap [6].    

The model Hamiltonian is a periodic version of $\frac{1}{r^2}$ exchange
given by
\begin{equation}
H = J \phi^2 \sum_{i<j} \frac{\vec{S}_{i}.\vec{S}_{j}}{\sin^2 [\phi (i-j)
]}
\end{equation}
where $\phi = \frac{\pi}{N}$ with $N$ the number of lattice spins with
lattice constant $1$. We consider only the antiferromagnetic case $J=1$.

In this paper we consider the $S=1$ Haldane-Shastry model (HS) and study
the ground-state properties. We use a modified Lanczos method [7,8]
to obtain the ground-state and first excited state energies and the
correlation functions for small systems of size up to $16$ spins.
We compare our results to those previously obtained for the Heisenberg
case stressing the similarities and differences. We estimate the
large system size limit of
these quantities using the extrapolation method of Vanden Broeck
and Schwartz (VBS) [9-11].

The exact energy spectrum has been obtained for the $S=\frac{1}{2}$ case
[1-4]. The groundstate is a spin singlet with energy [2]
\begin{equation}
\frac{E}{N J} = -\frac{\pi^2}{24} \left(1+ \frac{5}{N^2} \right)
\end{equation}
with a $\frac{1}{N^2}$ correction like for the Heisenberg model [11]. The
first excited state for $N$ even and for the Heisenberg model is a $S_T=1$ 
triplet where $S_T$ is the total spin of a $N$-spin state. 
In the case of the HS model the first excited state is however a
$S_T=1$ quartet. (For $N$ odd the groundstate is a quartet for both models).
The full spectrum has been obtained [4] showing a supermultiplet structure
such that the energies in units of $\frac{1}{4} \left(\frac{\pi}{N} \right)^2$
are integers [1]. A set of states with $N_{sp}$ spin-$\frac{1}{2}$ spinon
excitations with $S_T=S_T^z=\frac{N_{sp}}{2}$ (with $M=(N-N_{sp})/2$ reversed
spins) generates the full energy spectrum [4].

The Hamiltonian eq. (1) can be exactly diagonalized if we consider small
chains [1]. For spin-$S$ we have a basis of $(2S+1)^N$ states which grows
very fast as $S$ and $N$ grow. Taking $N \leq 8$ we have obtained the full 
spectrum for $S=1$. For $N$ even, the groundstate is again a singlet and the
first excited state is a $S_T=1$ triplet, as for the Heisenberg case. For $N$
odd the groundstate is a singlet in both models.

Since the size of the matrices to diagonalize increases exponentially with
the size of the chain, we used the modified Lanczos method [7] to obtain
the groundstate and the first excited state energies, together with the
groundstate correlation functions.

The size of the vectors can be considerably reduced using the symmetries
of the problem. The Hamiltonian commutes with the total spin operators
${(\vec{S}_{T})}^2$ and $S^{z}_{T}$, 
with the translation operator $T$,
the spin flip operator $R$ and the reflection operator $L$
$(i\rightarrow N+1-i, i=1,\ldots,N)$. The ground state has total
$S^{z}_{T}=0$ and one of the (degenerate) first excited states also has
$S^{z}_{T}=0$. 
One can then immediately reduce the states under consideration to this
subspace only. 

The operators $T$ and $L$ commute with $R$, i.e. one has $[T,R]=[L,R]=0$,
but they do not commute with each other, since $LT={T}^{-1}L$. The ground
state eigenvalues are $r=l=1$ and $t=1$, corresponding to an wavevector
$k=0$. The first excited state has $r=l=-1$ and $t=-1$, corresponding to
$k=\pi$ (for an even number of lattice points). 

Although in general the operator $L$ mixes $k$ and $-k$ states, the ground
state and the first excited state under consideration are simultaneous
eigenvectors of the operators $T,R$ and $L$. 

The action of the local operators $S^{\pm}_{i}$ and $S^{z}_{i}$ is simply
given in the direct product basis $|m^{z}_{1}>\cdots|m^{z}_{N}>$. In
general, these states are not eigenvectors of these additional symmetries.
One forms then the classes of states which are closed under them. One
starts with a state $|a>$, applies the translation operator $T$ $n_{t}$
times until one finds $T^{n_{t}}|a>=|a>$, where $n_{t}$ is necessarily a
divisor of $N$. One proceeds similarly with the operators $R$ and $L$ for
which $n _{r}=1,2$ and $n_{l}=1,2$. The multiplicity of this class is
$n_{t}n_{r}n_{l}$. 
The state $|a>$ is then the
representative of this class. The other classes are formed proceeding in
the same manner starting with other states, not already used, until the
state space is exhausted. An eigenvector of the operator $T$ (with
eigenvalue $t$ obeying to $t^{n_{t}}=1$) is obtained applying 
to its representatives the
symmetrization operator 
$\Omega^{\prime}_{t}=(1+t^{-1}T+(t^{-1}T)^{2}+\cdots+(t^{-1}T)^{n_{t}-1})$.
For the states in this class we have $\Omega_t^{\prime}=\frac{1}{s_t(a)} 
\Omega_t$, where the general operator 
$\Omega_{t} =
(1+t^{-1}T+(t^{-1}T)^{2}+\cdots+(t^{-1}T)^{N-1})$ 
and $s_t(a)=N/n_t$ is the symmetry factor.
The same applies to the operators $R$ and $L$. As stated above, when
$t=\pm 1$ it is possible to construct a simultaneous eigenvector of the
operators $T,R$ and $L$. In the cases in which the operator $R$ or $L$
does not introduce new states ($n_{r}=1$ or $n_{r}=1$) one should verify
that the eigenvalues induced by the first operator(s) are compatible with
those of the other operator(s). This allows to reduce even further the
number of classes. Any eigenvector with these eigenvalues is completely
defined by its projections on the representatives of the classes. 

The method goes as follows [7]. To obtain an approximate ground state wave 
function
we choose a trial wave function $\psi_0$ that
can not be orthogonal to the true ground-state. We define a state $\psi_1$
as [7]
\begin{equation}
\psi_1 = \frac{ \hat{H} \psi_0 - <H> \psi_0}{\left( <H^2>-<H>^2 \right)^{1/2}}
\end{equation}
where $<\psi_1 |\psi_1 >=1$, $<\psi_1 | \psi_0>=0$ and 
$<H^n>=<\psi_0 | \hat{H}^n |\psi_0>$. Defining a matrix of the Hamiltonian
in the basis $\psi_0,\psi_1$ we can diagonalize it [7] obtaining a next 
order approximation for the energy $\epsilon_1$, and ground state wave-function
 $\tilde{\psi}_0$, with
\begin{eqnarray}
\epsilon_1 & = & <H> + b \alpha \\
\tilde{\psi}_0 & = & \frac{\psi_0 + \alpha \psi_1 }{\left( 1+\alpha^2 
\right)^{1/2}}
\end{eqnarray}
where
\begin{eqnarray}
b & = & \left( <H^2>-<H>^2 \right)^{1/2} \\
\alpha & = & f-\left( f^2 +1 \right)^{1/2}
\end{eqnarray}
and
\begin{equation}
f = \frac{<H^3> - 3<H> <H^2> + 2 <H>^3}{2 \left(<H^2> - <H>^2 \right)^{3/2}}.
\end{equation}
Taking $\tilde{\psi}_0$ as the new $\psi_0$ we can iterate the method
to obtain a better estimate for the energy and ground state wave function. 

For the first excited state we proceed in a similar manner, starting from
a trivial wave function orthogonal to the true ground state (as guaranteed
by some symmetry argument), but not orthogonal to the true first excited
state. For small chains, where the use of the symmetries is not strictly
necessary, one can look for the excited state with ${S}^{z}_{T}=1$,
which is necessarily orthogonal to the ground state. However, for larger
chains one has to take into account the different symmetries, and with
the additional eigenvalues one can still look for the ${S}^{z}_{T}=0$
first excited state, as explained above. 

Since the complete Hamiltonian commutes with the operators $T,R$ and $L$
only transitions to the classes constructed above are allowed, even if
separate terms allow them. The action of the Hamiltonian on a state
is obtained writing, within each class, $H \Omega^{\prime} |a> = \frac {1}
{s(a)} \Omega H |a>$, where $\Omega$ is the product of the three symmetry
operators for $T$, $R$, and $L$. 
If $H|a> = \sum_b \alpha L^{p_{l}} R^{p_{r}} 
T^{p_{t}} |b>$
, where $p_{l,r,t}$ are integers and $|b>$ is the representative of a class,
one finally finds $ <b|H \Omega^{\prime} |a>=\sum_b \frac{s(b)}{s(a)} \alpha
l^{p_{l}} r^{p_{r}} t^{p_{t}}.$ One also has to take into 
account the multiplicity of the classes when normalizing states and making
inner products of states. 

The accuracy of the method was tested against the exact diagonalization of
the full spectrum for sizes up to $N=8$. We considered system sizes up to
$N\leq 16$ (with $N$ even).

This procedure gives the results for the several finite-size systems. We are
however interested in the infinite system limit and standard extrapolation
methods [9] have to be used like
the VBS method [9,10]. In this
method we want to estimate the limit of a finite sequence $P_n$ ($n=1,...,N$).
Defining
\begin{eqnarray}
P_n^{(m+1)} & = & P_n^{(m)} + \frac{1}{Q_n^{(m)} - Q_{n-1}^{(m)}} \\
Q_n^{(m)} & = & \alpha_m Q_n^{(m-1)} + \frac{1}{P_{n+1}^{(m)} - P_n^{(m)}}
\end{eqnarray}
where $Q_n^{(-1)}=0$, $P_n^{(0)}=P_n$ we obtain an estimate of the sequence
iterating (like the recursive rule due to Wynn). If $\alpha_m=0$ 
this is the Aitken-Shanks 
transformation which is adequate for exponential behavior. To generate the 
Pad\'{e}-Shanks transformation we select $\alpha_m=1$. A power law behavior is 
well fitted 
choosing the Hamer and Barber's transformation $\alpha_m=-[1-(-1)^m]/2$.
We get an estimate of the asymptotic value of the
sequence $P_n$ [9,10] selecting $\alpha_m$ appropriately.

\bf
i) Ground-State Energy
\rm

In Table I we show the values for the groundstate energy of the $S=1$ 
Haldane-Shastry chain with sizes $N=4$ to $16$ ($N$ even). 
A linear fit of the groundstate energy per spin as a function of 
$\frac{1}{N^2}$
yields ($1.2568\pm 0.0016$). Using the 
more accurate VBS method
yields the results shown in Table II. 
These results indicate that the behavior is not purely $\frac{1}{N^2}$ in
contrast to the $S=\frac{1}{2}$ case. Other terms have to be included,
particularly for small values of $N$.
Since the spectrum has a gap (see below) we expect that the exponential 
behavior is more adequate [12,13]. It has been found for the Heisenberg 
model [13] that $\alpha_m=1$ yields the best results. This has been found 
looking at the decay length at size $N$ and requiring that 
\begin{equation}
\xi(n,m) = \frac{2}{\ln \left( \frac{P_{n-1}^{(m-1)}-P_{n}^{(m-1)}}{
P_{n}^{(m-1)}-P_{n+1}^{(m-1)}} \right)}
\end{equation}
be such that
\begin{equation}
\xi(n,m) < \xi(n+1,m-1).
\end{equation}
Only the Pad\'{e}-Shanks transformation yielded consistent results [13].  
In our case, however, all three cases yielded consistent tables and therefore
we include the three choices.

\bf
ii) Gap
\rm

We consider now the gap (the difference in energy between the first excited
state and the groundstate). In Table III we give the values of the gap
as a function of $N$ for the Haldane-Shastry model.
In Fig. 1 we show the gap and the gap per
spin 
as a function of $N$
for the two models. The results seem to indicate that the gap is
finite for both models in the $N\rightarrow \infty$ limit in agreement
with Haldane's conjecture [6] and with previous results for the Heisenberg
case. The VBS extrapolated values are shown in Table IV. 
In the case of the gap the Pad\'{e}-Shanks yields once again consistent results
but the other two do not obey eq. (12) for the last iteration of eqs. (9,10).
We recall that, for comparison, the value of the gap for the Heisenberg 
model is $0.41050$ [13,14].
The extrapolated value of the 
gap per
spin is close to but not zero due to the small number of points
considered. 

\newpage
\bf
iii) Correlation Functions
\rm

The groundstate correlation functions are defined by
\begin{equation}
C_m(N) = \frac{3}{N} \sum_{i=1}^{N} \frac{<S_i^z S_{i+m}^z>}{S(S+1)}
\end{equation}
In Table V we show them for the same set of system sizes
and in Table VI the VBS extrapolated values.

In Fig. 2 we compare the correlation functions with those for the
Heisenberg model. 
For a finite system the energy scale of the two models
is not the same. However, in the infinite system limit the nearest-neighbor 
interaction is the same. We feel therefore that it is worthwhile to compare 
the behavior for the two models. 
In Fig. 2a $|C_1|$ is displayed as a function
of $N$; in Fig.2b we show $C_2$ also as a function of $N$ and 
in Fig. 2c we show $C_m$ for $N=16$ as a function of $m$. The general
trend is that, in spite of the long-range nature of the interaction,
the correlation functions for the Haldane-Shastry model decay faster
(in the sense that the numerical values are smaller) than those for 
the Heisenberg model. A possible interpretation is that, 
similarly to the
$S=\frac{1}{2}$ case, 
there may be
an absence of interactions between the spinons. 

It has been argued that the behavior of the correlation functions
$C_{N/2}$ when $N \rightarrow \infty$ reflects their behavior with
distance in the infinite system [15,8]. If there is a gap in the
spectrum it is expected that
\begin{equation}
\lim_{m\rightarrow \infty} |C_m| \sim B e^{-\frac{m}{\xi}}
\end{equation}
instead of the power-law behavior observed in the $S=\frac{1}{2}$ case
(in both models). 
In Fig. 3 we plot $\ln |C_{N/2} |$ vs. $N$ for both models. 
A linear fit yields for the Haldane-Shastry model
\begin{equation}
\ln |C_{N/2} | = (-0.02 \pm 0.02) - (0.162 \pm 0.0019) N
\end{equation}
which corresponds to a correlation length of the order of $\xi=3.1$. A similar
analysis for the Heisenberg model gives $\xi_H$ of the order of $4.9$.
The quality of the fits is similar for both models. 
However, the correlation length for the Heisenberg model has been estimated
to be $\xi_H=6.03$ [14] obtained from a fit to the large distance limit of 
the Bessel function $K_0(r/\xi)$, which asymptotically behaves as $\sim
(\xi/r)^{1/2} e^{-r/\xi}$, and, therefore, a plot like in Fig. 3 underestimates
the correlation length for the Heisenberg model.

In Fig. 4a we show $\ln |C(m)|$ as a function of $m$ for
$N=16$. At large $m$ we expect an exponential decay of the form of eq. (14)
with possible corrections.
It is clear that the exponential behavior has not been fully 
reached. The exponential behavior is probably, as for the Heisenberg model,
only to leading order. 
We have tried several functional forms like those for the
Heisenberg model. 
In Fig. 4b we plot $F(m)=\ln (|C(m)| m^{1/2})$ as a function of $m$ for $N=16$.
It is clear that for these system sizes the fit is poor even for the Heisenberg
model.
One requires larger systems to reach the limiting
behavior. 
Note, however, that the functional form for the Haldane-Shastry has not been
determined and therefore the numerical results are not conclusive to extrapolate
to the large $N$ limit.
On the other hand,
the linear fit of Fig. 3 is quite good for both models. We take the value obtained 
for $\xi$ for the Haldane-Shastry model to be the estimate obtained by this fit.  
Larger system sizes and the appropriate functional form are therefore required to 
more accurately determine the
correlation length for the Haldane-Shastry model.

In summary, in this work we have studied the groundstate of the $S=1$
Haldane-Shastry model, which has been recently presented [1,2]
and solved exactly
for $S=\frac{1}{2}$. The importance of the model lies on the fact that it
belongs to a new class [5] of integrable systems. We have confirmed that,
according to arguments by Haldane, the model shows a finite gap in the
energy spectrum. We based our result on the extrapolated value of the 
calculated gap for various small chains and on the exponential behavior of the
correlation function $C_{N/2}$ with $N$. We have also obtained, by comparison
to the Heisenberg case, that the correlation functions for the
Haldane-Shastry model decay faster as a function of both $m$ and $N$. 
The same happens in the $S=1/2$ case where the spectrum is gapless. This is
due to the oscillatory nature of the (positive) interaction in this model.
The estimated correlation length is smaller for the Haldane-Shastry model
and the value of the gap is larger than for the Heisenberg model. 
These exact
numerical results further extend the similarities and differences between
the two models for a case ($S=1$) which is not exactly solvable, in
contrast to the $S=\frac{1}{2}$ case.

\vspace{\baselineskip}
\bf
\centerline{Acknowledgements}
\rm
One of us (PDS) acknowledges partial support in the
form of a PRAXIS XXI Fellowship and a helpfull discussion with Z. Tesanovic
and A. Nersesyan.

\vspace{\baselineskip}
\vspace{\baselineskip}
$^{*}$ On leave of absence from Departamento de F\'{\i}sica, Instituto
Superior T\'{e}cnico, Av. Rovisco Pais, 1096 Lisboa Codex, Portugal.

\newpage
\vspace{\baselineskip}
\bf
\centerline{REFERENCES}
\rm

\vspace{\baselineskip}
\noindent [1] Haldane, F.D.M. : Phys. Rev. Lett. {\bf 60}, 635 (1988).

\vspace{\baselineskip}
\noindent [2] Shastry, B.S. : Phys. Rev. Lett. {\bf 60}, 639 (1988).

\vspace{\baselineskip}
\noindent [3] Gebhard, F., Vollhardt, D. : Phys. Rev. Lett. {\bf 59}, 
1472 (1987).

\vspace{\baselineskip}
\noindent [4] Haldane, F.D.M. : Phys. Rev. Lett. {\bf 66}, 1529 (1991).

\vspace{\baselineskip}
\noindent [5] Haldane, F.D.M. : {\em Correlation Effects in Low-Dimensional
Electron Systems}, 3, ed. Okiji, A., Kawakami, N., Springer-Verlag 1994.

\vspace{\baselineskip}
\noindent [6] Haldane, F.D.M. : Phys. Lett. {\bf 93}A, 464 (1983); 
Phys. Rev. Lett. {\bf 50}, 1153 (1983).

\vspace{\baselineskip}
\noindent [7] Gagliano, E.R., Dagotto, E., Moreo, A., Alcaraz, F.C. : 
Phys. Rev. B {\bf 34}, 1677 (1986). 

\vspace{\baselineskip}
\noindent [8] Moreo, A. : Phys. Rev. B {\bf 35}, 8562 (1987). 

\vspace{\baselineskip}
\noindent [9] Barber, M.N. : {\em Phase Transitions}, vol. 8, 146 ed.  
Domb, C., Lebowitz, J., New York, Academic 1983. 

\vspace{\baselineskip}
\noindent [10] Betsuyaku, H.: Phys. Rev. B {\bf 34}, 8125 (1986). 

\vspace{\baselineskip}
\noindent [11] Bonner, J.C., Fisher, M.E. : Phys. Rev. {\bf 135}, A640
(1964). 

\vspace{\baselineskip}
\noindent [12] Golinelli, O., Jolicouer, Th., Lacaze, R. : Phys. Rev. B 
{\bf 45},
9798 (1992). 

\vspace{\baselineskip}
\noindent [13] Golinelli, O., Jolicouer, Th., Lacaze, R.: Phys. Rev. B {\bf 50}
,
3037 (1994). 

\vspace{\baselineskip}
\noindent [14] White, S.R., Huse, D.A. : Phys. Rev. B {\bf 48}, 3844 (1993).

\vspace{\baselineskip}
\noindent [15] Botet, R., Jullien, R., Kolb, M. : Phys. Rev. B {\bf 28}, 
3914 (1983). 

\vspace{\baselineskip}
\noindent [16] See ref. 13 or Liang, S. : Phys. Rev. Lett. {\bf 64}, 1597 (1990). 

\newpage 
TABLE I- Groundstate energy and groundstate energy per spin
as a function of $N$.  

\vspace{\baselineskip}
$
\begin{array}{ccc}
N & -E_N & -E_N/N \\
  &      &        \\
4 & 6.168503 & 1.542126 \\
6 & 8.270682 & 1.378447 \\
8 & 10.59553 & 1.324441 \\
10 & 13.00953 & 1.300953 \\
12 & 15.46690 & 1.288908  \\
14 & 17.94784 & 1.281988  \\
16 & 20.44263 & 1.277664
\end{array}
$

\vspace{\baselineskip}
\vspace{\baselineskip}
Table II- Extrapolated values for $-E_N/N$ for the Hamer-Barber (H-B), 
Pad\'{e}-Shanks (P-S) and Aitken-Shanks (A-S) transformations eqs. (9,10).

\vspace{\baselineskip}
$
\begin{array}{ccc}
H-B & P-S & A-S \\
  &      &        \\
1.263147 & 1.267894 & 1.265328 
\end{array}
$

\newpage
TABLE III- Gap between the groundstate and the first excited state
as a function of $N$. 

\vspace{\baselineskip}
$
\begin{array}{ccc}
N & Gap & Gap/N \\
  &     &  \\
4 & 1.2337 & 0.30843 \\
6 & 0.87226 & 0.14538 \\
8 & 0.71962 & 0.08995 \\
10 & 0.64551 & 0.06455 \\
12 & 0.60678 & 0.05057 \\
14 & 0.58550 & 0.04182 \\
16 & 0.57333 & 0.03583 
\end{array}
$

\vspace{\baselineskip}
\vspace{\baselineskip}
Table IV- Extrapolated values for the gap for the Hamer-Barber (H-B), 
Pad\'{e}-Shanks (P-S) and Aitken-Shanks (A-S) transformations eqs. (9,10).

\vspace{\baselineskip}
$
\begin{array}{ccc}
H-B & P-S & A-S \\
  &      &        \\
0.55045 & 0.55439 & 0.55285 
\end{array}
$

\newpage
TABLE V- Correlation functions $C_m$ for the set of values $N=4$
to $16$ with $N$ even.

\vspace{\baselineskip}
$
\begin{array}{rrrrrrrrr}
& N & 4 & 6 & 8 & 10 & 12 & 14 & 16  \\
m &  &  &   &   &    &    &    &  \\
1 & & -0.75000 & -0.71660 & -0.70630 & -0.70210 & -0.70017 & -0.69923 & 
-0.69876 \\
2 & & 0.50000 & 0.40744 & 0.37357 & 0.35722 & 0.34872 & 0.34420 & 0.34177 \\
3 & &         & -0.38168 & -0.29832 & -0.26270 & -0.24504 & -0.23586 &
-0.23099 \\
4 & &         &          & 0.26209  & 0.20479 & 0.17697 & 0.16261 & 0.15501 \\
5 & &         &          &          & -0.19443 & -0.14946 & -0.12729 & 
-0.11582  \\
6 & &         &          &          &        & 0.13797  & 0.10601 & 0.08972 \\ 
7 & &         &          &          &        &  & -0.10088 & -0.07720 \\ 
8 & &         &          &          &        &  & & 0.07251 
\end{array}
$

\newpage
TABLE VI- Extrapolated values for the correlation functions $C_m$ for the 
same set of values of $N$ using  
the Hamer-Barber (H-B), Pad\'{e}-Shanks (P-S) and Aitken-Shanks (A-S) 
transformations eqs. (9,10).

\vspace{\baselineskip}
$
\begin{array}{rrrr}
m & H-B & P-S & A-S  \\
  &     &     &      \\
1 & -0.69829 & -0.69828 & -0.69827 \\
2 & 0.33890 & 0.33887   & 0.33893  \\
3 & -0.22531 & -0.22559 & -0.22546 \\
  & -0.22533 & -0.22536 & -0.22534 \\
4 & 0.14616  & 0.14624  & 0.14620 \\
5 & -0.10573 & -0.10573 & -0.10573 \\
  & -0.10352 & -0.10352 & -0.10352 \\
6 & 0.07279  & 0.07279  & 0.07279
\end{array}
$

\newpage

Fig. 1- Gap and gap per spin for both models, Heisenberg (H) and
Haldane-Shastry (HS), as a function of $N$.

\vspace{\baselineskip}
Fig. 2- Correlation functions for the Heisenberg (H) and Haldane-Shastry
(HS) models for a) $|C_1|$, b) $C_2$ as a function of $N$ and c) $C_m$ for
$N=16$ vs. $m$.

\vspace{\baselineskip}
Fig. 3- The logarithm of the correlation function $C_{N/2}$ as a function of $N$ for the
Heisenberg (H) and Haldane-Shastry (HS) models. The solid line is the
linear fit showing the exponential behavior.

\vspace{\baselineskip}
Fig. 4- a) The logarithm of the correlation functions $|C(m)|$ and b) 
$F(m)=\ln (|C(m)| m^{1/2})$ as a function
of $m$ for $N=16$ for the Heisenberg (H) and Haldane-Shastry (HS) models.

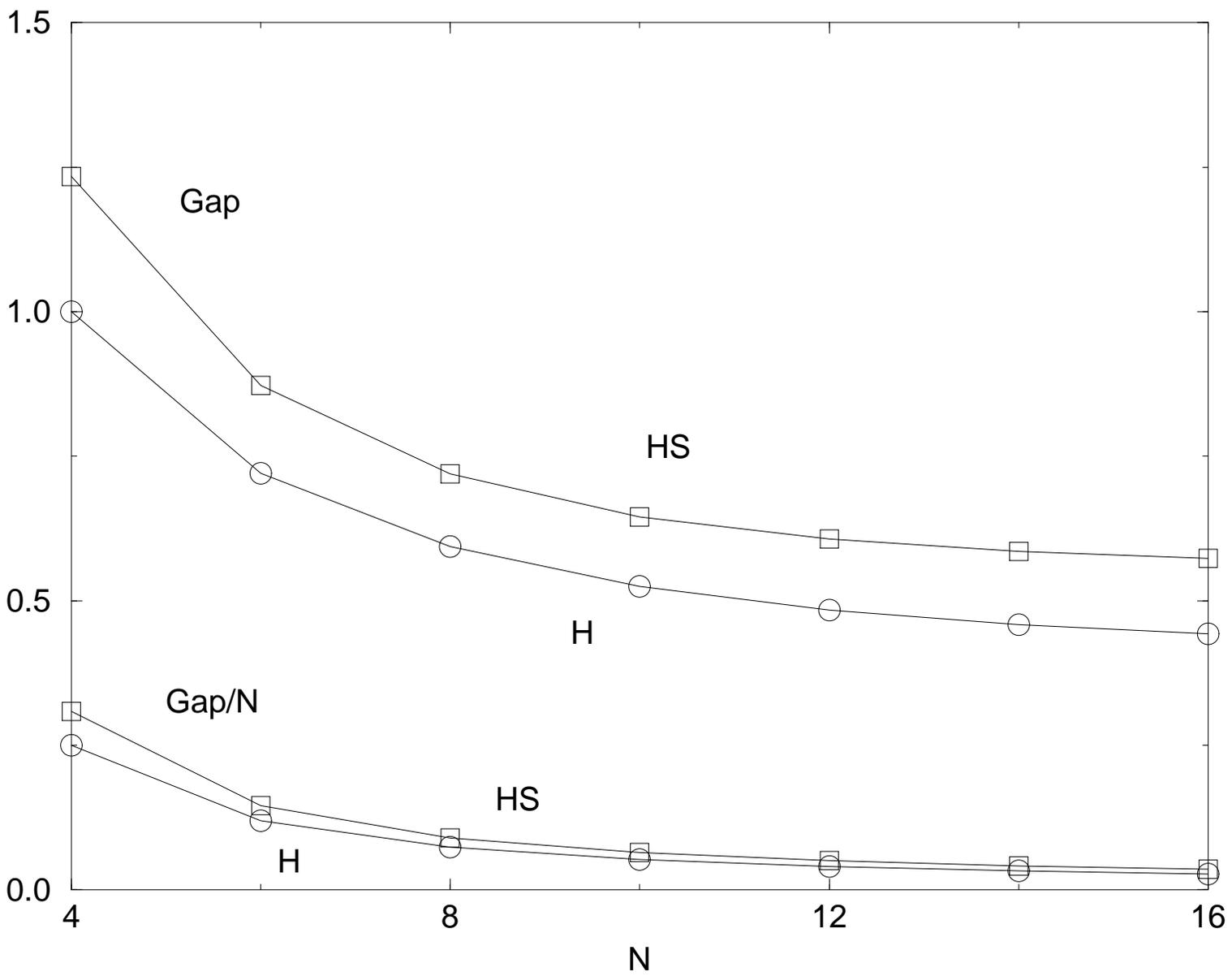
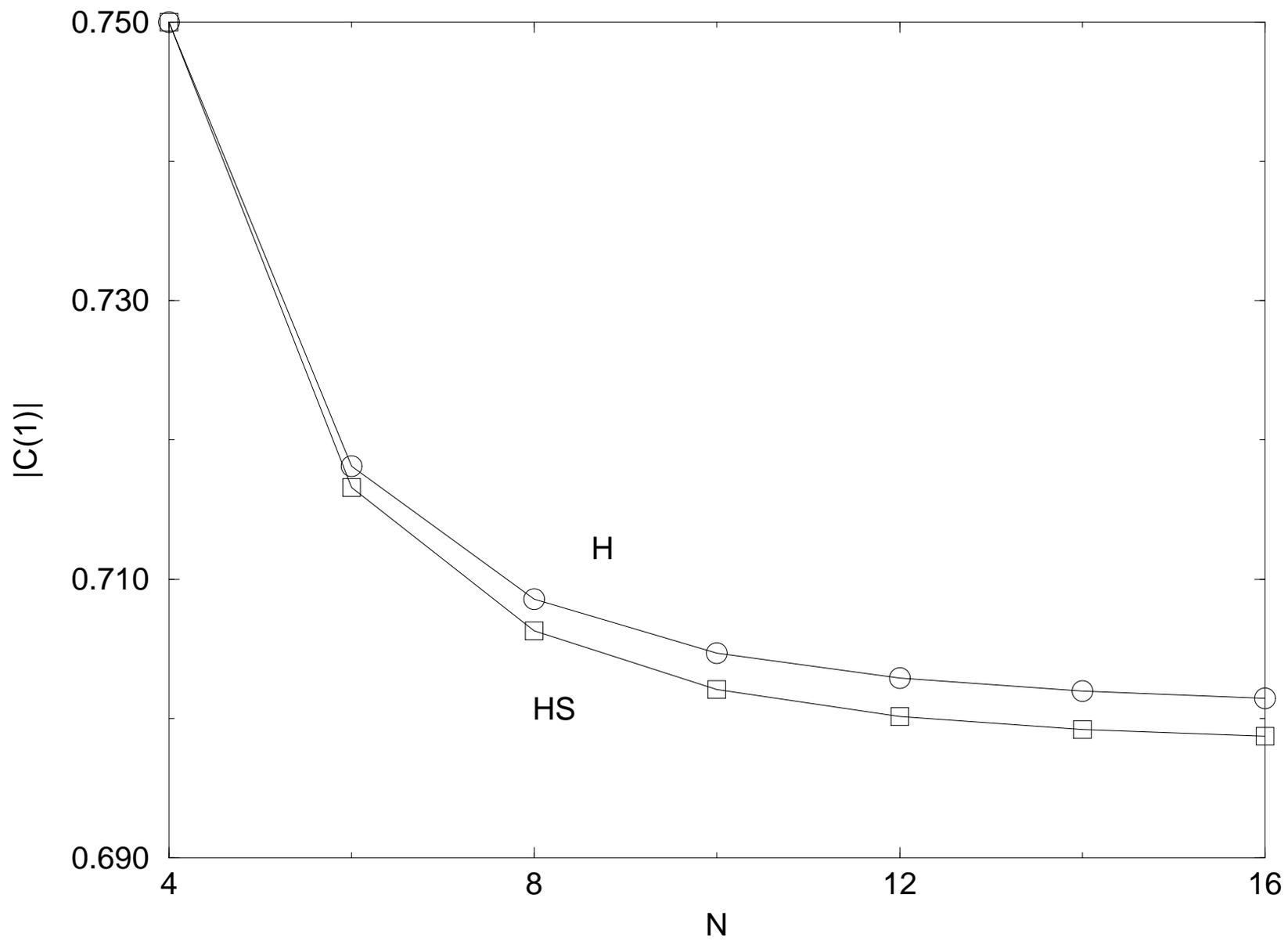
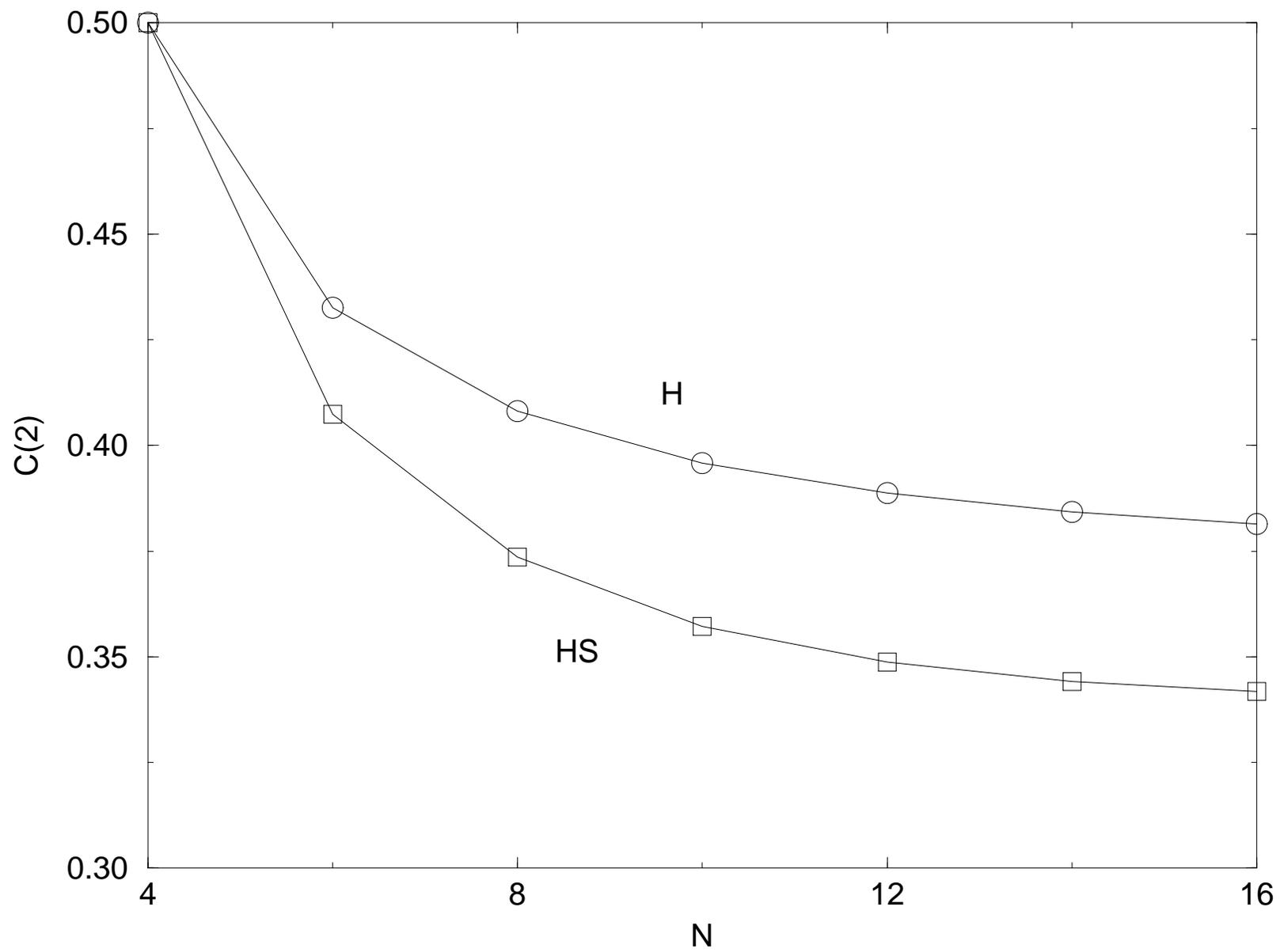
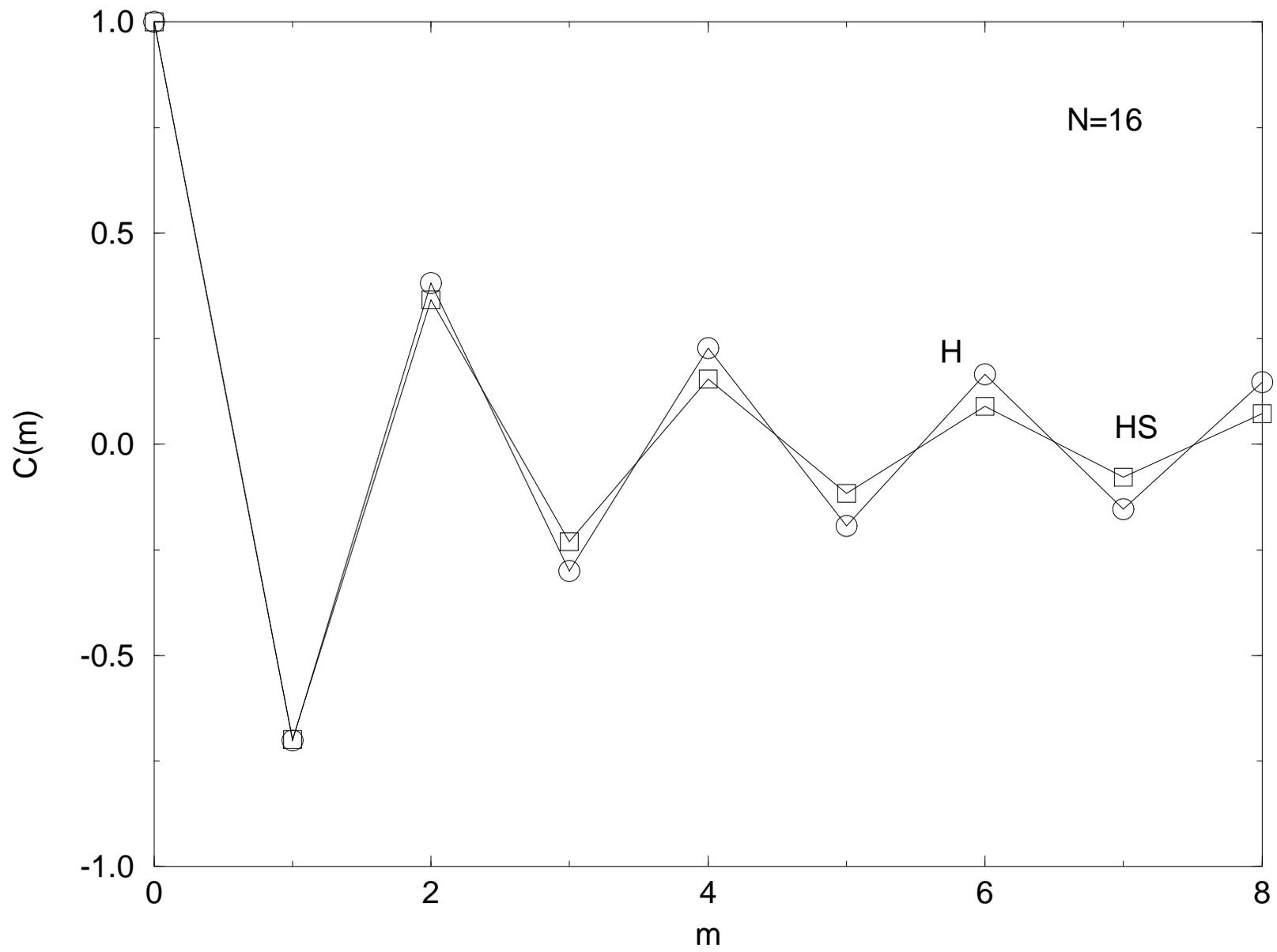
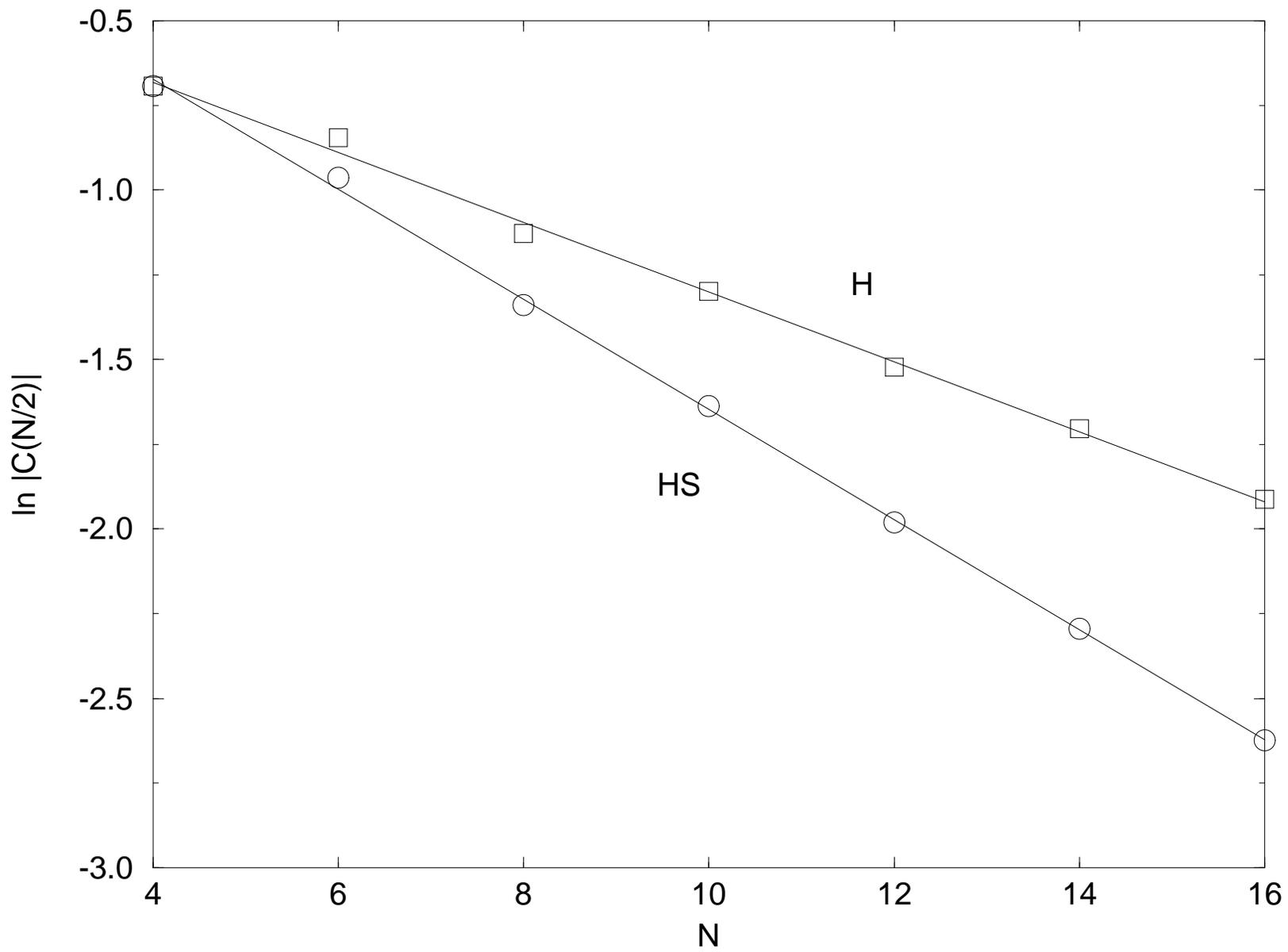
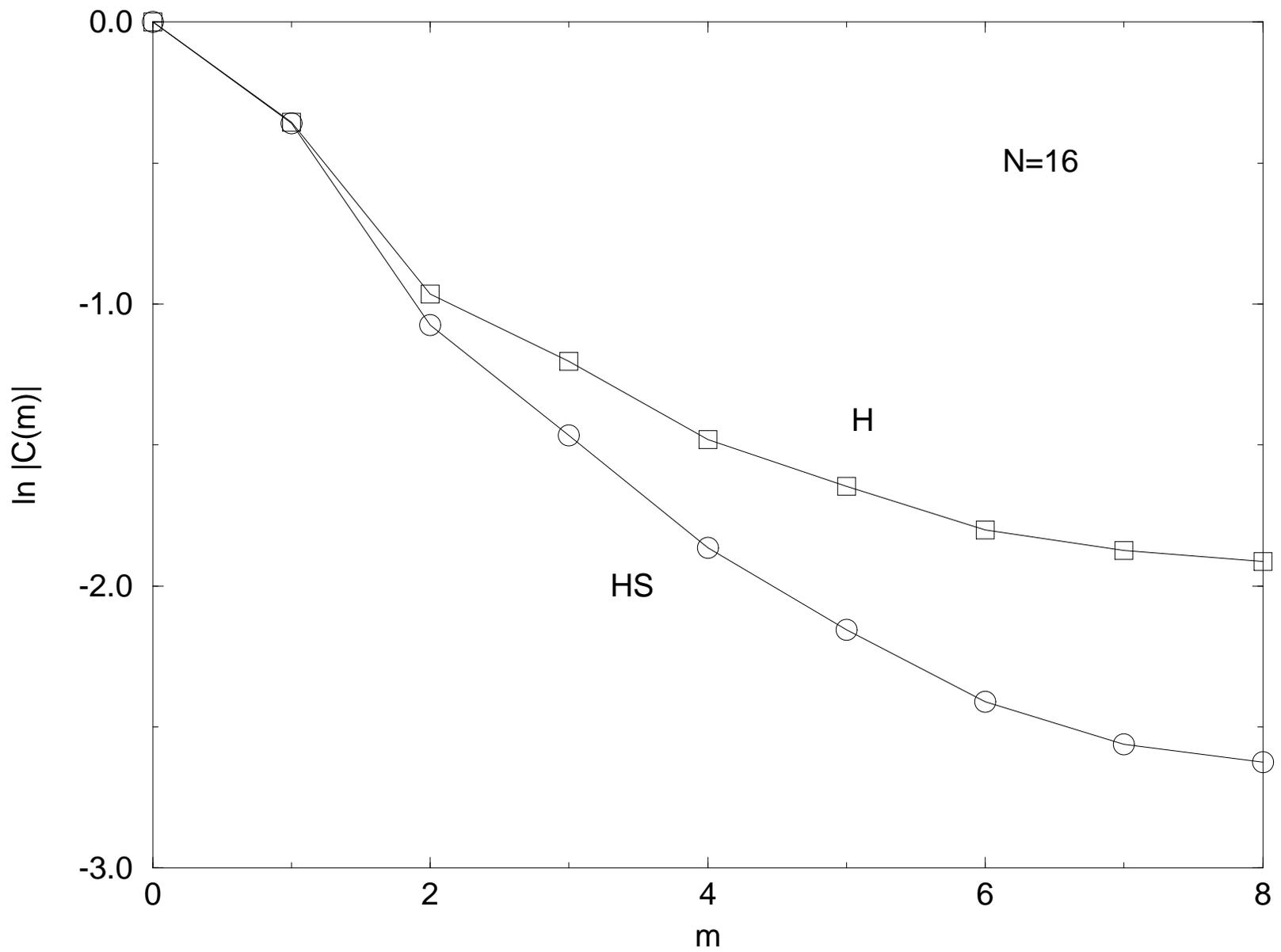
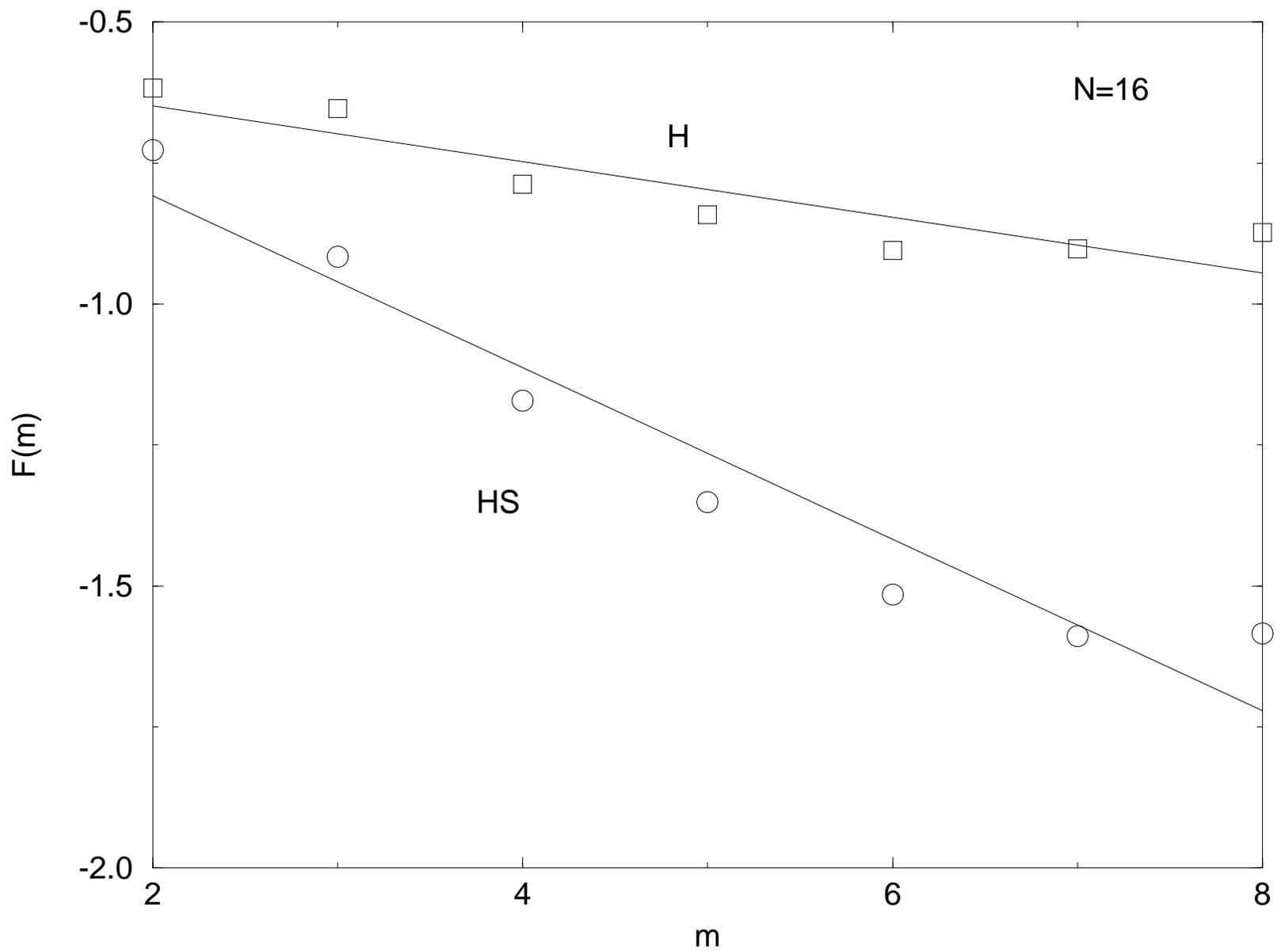
\end{document}